\documentclass[conference]{IEEEtran}
\IEEEoverridecommandlockouts
\usepackage{cite}
\usepackage{amsmath,amssymb,amsfonts}
\usepackage{algorithmic}
\usepackage{graphicx}
\usepackage{textcomp}
\usepackage{xcolor}
\usepackage[acronym]{glossaries}
\usepackage{multirow}
\usepackage[font=small]{caption}
\usepackage{booktabs}
\usepackage{subcaption}

\def\BibTeX{{\rm B\kern-.05em{\sc i\kern-.025em b}\kern-.08em
    T\kern-.1667em\lower.7ex\hbox{E}\kern-.125emX}}

\begin{document}

% \title{O-POPE: XXX}
\title{O-POPE: High-Frequency Pipelined Outer Product based GEMM acceleration with minimal buffering overhead}

%%%%%%%%%%%%%%%%%%%%%%%%%%%%%%%%%%
%%            AUTHORS           %%
%%%%%%%%%%%%%%%%%%%%%%%%%%%%%%%%%%

\author{
  Danilo Cammarata\textsuperscript{1},
  Angelo Garofalo\textsuperscript{2},
  Luca Benini\textsuperscript{1,2} \\
  \textsuperscript{1}\textit{ETH Zurich, Zurich, Switzerland},
  \textsuperscript{2}\textit{Università di Bologna, Bologna, Italy} \\
  \{dcammarata, lbenini\}@iis.ee.ethz.ch, angelo.garofalo@unibo.it
}
% \author{Authors hidden for double-blind review}

%%%%%%%%%%%%%%%%%%%%%%%%%%%%%%%%%%
%%          PAPER FILES         %%
%%%%%%%%%%%%%%%%%%%%%%%%%%%%%%%%%%

\maketitle

\newcommand\red[1]{\textcolor{red}{#1}\PackageWarning{}{#1!}}

\newcommand{\MaxUtilQuad}     {{99.9\%}}
\newcommand{\MaxUtilEight}    {{99.4\%}}
\newcommand{\MaxUtilSixTeen}  {{96.9\%}}
\newcommand{\MaxUtil}         {{99.97\%}}
\newcommand{\PerfGain}        {{33\%}}
\newcommand{\AreaGain}        {{9\%}}
\newcommand{\PowerGain}       {{8\%}}
\newcommand{\MaxFreq}         {{890}}
\newcommand{\OpFreq}          {{700}}
\newcommand{\AxonFreq}        {{550}}
\newcommand{\OpenGeMMFreq}    {{200}}

\newacronym{au}{AU}{arithmetic unit}
\newacronym{ce}{CE}{compute element}
\newacronym{ci}{CI}{confidence interval}
\newacronym{cnn}{CNN}{convolutional neural network}
\newacronym{dma}{DMA}{direct memory access}
\newacronym{dmr}{DMR}{dual modular redundancy}
\newacronym{dtr}{DTR}{dual temporal redundancy}
\newacronym{due}{DUE}{detectable unrecoverable error}
\newacronym{dut}{DUT}{device under test}
\newacronym{ecc}{ECC}{error correction codes}
\newacronym{ff}{FF}{flip-flop}
\newacronym[firstplural={first-in, first-out (FIFO) buffers},longplural={first-in, first-out buffers}, first=AA]{fifo}{FIFO}{first-in, first-out}
\newacronym{fma}{FMA}{floating multiply-add}
\newacronym{fp}{FP}{floating point}
\newacronym{pe}{PE}{Processing Element}
\newacronym{fpga}{FPGA}{field-programmable gate array}
\newacronym{fpu}{FPU}{floating-point unit}
\newacronym{fsm}{FSM}{finite-state machine}
\newacronym{ge}{GE}{gate equivalent}
\newacronym{gemm}{GEMM}{General matrix multiply}
\newacronym{gpio}{GPIO}{general purpose input/output}
\newacronym{gpu}{GPU}{Graphical Processing Unit}
\newacronym{hci}{HCI}{Heterogeneous Cluster Interconnect}
\newacronym{hpc}{HPC}{high-performance computing}
\newacronym{hwpe}{HWPE}{Hardware Processing Engine}
\newacronym{is}{IS}{input stationary}
\newacronym{isa}{ISA}{instruction set architecture}
\newacronym{lsu}{LSU}{Load-Store Unit}
\newacronym{mac}{MAC}{multiply–accumulate}
\newacronym{matmul}{MatMul}{matrix multiplication}
\newacronym{mftf}{MFTF}{mean fluence to failure}
\newacronym{ml}{ML}{machine learning}
\newacronym{mrf}{MRF}{matrix register file}
\newacronym{mttf}{MTTF}{mean time to failure}
\newacronym{npu}{NPU}{Neural Processing Unit}
\newacronym{odrg}{ODRG}{on-demand redundancy grouping}
\newacronym{os}{OS}{output stationary}
\newacronym{pcb}{PCB}{printed circuit board}
\newacronym{pdk}{PDK}{process design kit}
\newacronym{ppa}{PPA}{power, performance, and area}
\newacronym{pulp}{PULP}{parallel ultra-low power}
\newacronym{rf}{RF}{register file}
\newacronym{rhbd}{RHBD}{radiation hardened by design}
\newacronym{rpi}{RPi}{Raspberry Pi}
\newacronym{rvv}{RVV}{RISC-V Vector}
\newacronym{sa}{SA}{systolic array}
\newacronym{sdc}{SDC}{silent data corruption}
\newacronym{sdotp}{SDOTP}{sum of dot product}
\newacronym{secded}{SECDED}{single error correction, double error detection}
\newacronym{see}{SEE}{single-event effect}
\newacronym{sefi}{SEFI}{single-event functional interrupt}
\newacronym{sel}{SEL}{single-event latchup}
\newacronym{set}{SET}{single-event transient}
\newacronym{seu}{SEU}{single-event upset}
\newacronym{soc}{SoC}{system-on-chip}
\newacronym{sram}{SRAM}{static random-access memory}
\newacronym{tcdm}{TCDM}{tightly coupled data memory}
\newacronym{tcls}{TCLS}{triple-core lockstep}
\newacronym{tid}{TID}{total ionizing dose}
\newacronym{tmr}{TMR}{triple modular redundancy}
\newacronym{tpu}{TPU}{Tensor Processing Unit}
\newacronym{udma}{$\mu$DMA}{I/O DMA}
\newacronym{vrf}{VRF}{vector register file}
\newacronym{wdt}{WDT}{watchdog timer}
\newacronym{ws}{WS}{weight stationary}
\newacronym{xif}{XIF}{OpenHW Group CORE-V-X interface}

\begin{abstract}
\gls{gemm} dominates both execution time and energy consumption of modern \gls{ml} workloads, placing increasing pressure on hardware efficiency.
While quantization mitigates computational and data movement costs, accuracy-sensitive tasks such as training still require higher-precision floating-point formats. 
Existing floating-point \gls{gemm} accelerators face trade-offs between operating frequency, arithmetic utilization, and buffering overhead.
This work presents O-POPE, a scalable outer-product engine that achieves concurrently high utilization, low overhead, and a fast operating frequency by repurposing \gls{fpu} pipeline registers as buffers. This solution leverages the data-reuse advantages of output-stationary outer-product execution and enables 1 GHz (0.72 V) operation in 12 nm FINFET technology with less than 2\% buffer area for a 2048-MACs configuration.
Our evaluation shows that O-POPE achieves up to {\MaxUtil} \gls{fpu} utilization and improves performance (1.33$\times$), performance density by {\AreaGain}, and energy efficiency by {\PowerGain}, compared to state-of-the-art floating-point GEMM accelerators.
\end{abstract}


\begin{figure*}[h]
    \centering
    \includegraphics[width=0.9\textwidth]{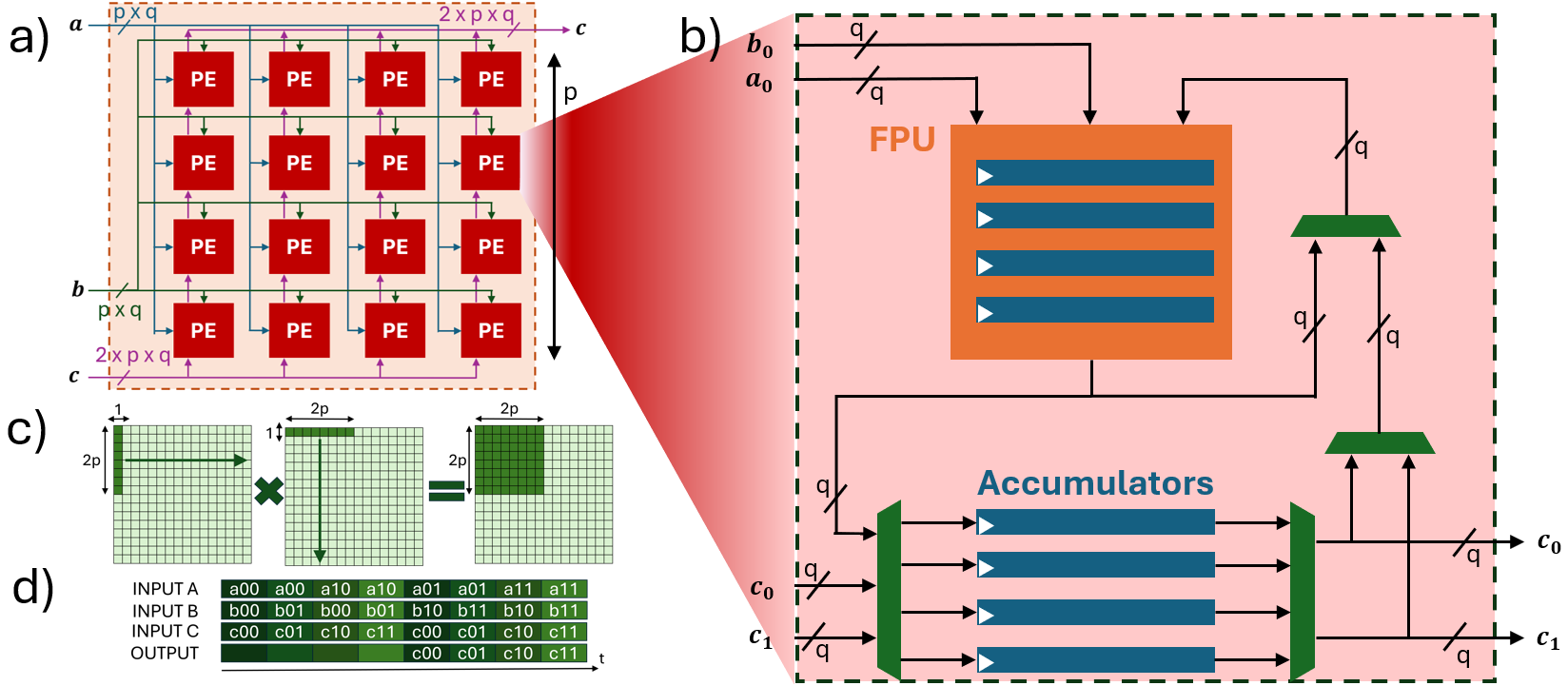}
    \caption{(a) O-POPE architecture; (b) O-POPE \acrlong{pe}; (c) GEMM execution flow; (d) FPU inputs/outputs across multiple cycles.}
    \label{fig:architecture}
\end{figure*}
\section{Introduction} 
\label{sec:01_introduction}

Recent years have seen a surge in the adoption of ML applications across several domains, thanks to their capabilities, efficiency, and accessibility~\cite{ML_adv}. 
The operator that dominates execution time and energy consumption in ML workloads is the \gls{gemm}, which accounts for 42.8\%--96.6\% of runtime in transformer-based models~\cite{gemm_impact}.
The rapid growth in model size and complexity further amplifies this computational demand~\cite{scaling_laws}, raising critical concerns about execution time and energy efficiency.
Thus, researchers actively investigate approaches aimed at curbing \gls{gemm} cost, including precision reduction through quantization~\cite{Quantization}, matrix \gls{isa} extensions~\cite{MatrixISA}, and specialized architectures~\cite{Arch}.

Quantization reduces data transfer and storage costs~\cite{fp4,ocpmx,nvfp4} but accuracy-sensitive workloads such as ML training still require 8-bit or higher-precision floating-point formats~\cite{quantized_fp8,fp64}. Hence, efficient floating-point \gls{gemm} acceleration remains essential.

Traditional scalar and vector \gls{isa}s incur costly register-file accesses and limited data reuse due to the lack of two-dimensional parallelism~\cite{Quad}. Existing matrix extensions ~\cite{arm,Intel,IBM} and emerging RISC-V proposals~\cite{ime,ame} address this limitation but introduce execution-time and energy overhead, motivating specialized \glspl{npu}~\cite{npu}.

Among these application-specific architectures, several floating-point \gls{gemm} accelerators have emerged, spanning a range of microarchitectural choices that affect both execution time and energy cost.
Given the importance of reducing execution time and energy cost in modern ML workloads, floating-point \gls{gemm} accelerators require: (i) minimal buffer and control overhead, (ii) high arithmetic utilization, and (iii) high operating frequency while maintaining scalability.

Existing architectures optimize at most two of these key performance metrics.
Inner product \gls{gemm} accelerators, such as RedMulE~\cite{redmule}, can sustain high operating frequencies and utilization. 
However, they incur substantial buffering overhead due to their input stationary dataflow. 
In these architectures, input buffer capacity has to scale linearly with the number of \glspl{fpu} and pipeline stages. 
In contrast, outer-product \gls{gemm} accelerators provide more favorable scaling properties, as input buffer capacity only grows with the square root of the number of \glspl{fpu} and internal reuse of data transferred from the memory system is maximized.
% As shown in Section II, an output stationary dataflow based on outer product operations minimizes memory transfers and buffering overhead for \gls{gemm}s.
Despite these advantages, existing outer-product \gls{gemm} accelerators do not achieve high operating frequencies. 
For instance, accelerators based on \texttt{FP16} \gls{mac} units from the FPnew floating-point unit~\cite{fpnew} achieve 555\,MHz in 22\,nm node (Sauria~\cite{sauria2023}), 550\,MHz in 12\,nm node (Gemmini~\cite{gemmini}), and 550\,MHz in 7\,nm node (Axon~\cite{axon}).
This frequency bottleneck arises from limited pipelining in the \glspl{fpu}. 
Pipeline registers increase the latency of floating-point operations and then profoundly affect the dataflow, as they can introduce significant area and energy overhead for buffering to meet the tight synchronization constraints imposed by systolic architectures.

The key intuition in this work is that we can exploit floating-point pipeline registers as implicit buffers. Thus, high-frequency outer-product execution becomes possible without introducing additional buffers, while achieving extremely high utilization.
Leveraging this insight, we present O-POPE\footnote{https://github.com/pulp-platform/opope}, a scalable outer-product engine with near-ideal utilization for FP \gls{gemm}s.
O-POPE implements an output-stationary outer-product dataflow on a semi-systolic architecture, as defined in ~\cite{htkung}, supporting configuration for multiple floating-point datatypes (\texttt{FP8}, \texttt{FP16}, \texttt{FP32}) and straightforward adaptability to integer datatypes. 
O-POPE exploits the pipeline registers of its \glspl{fpu} for both frequency scaling and buffering, achieving up to 99.97\% utilization. Moreover, O-POPE operates at 1\,GHz at 0.72\,V on a 12\,nm node across multiple semi-systolic array sizes ($4 \times 4$, $8 \times 8$, $16 \times 16$, $32 \times 32$), without significant buffer area as in existing inner product \gls{gemm} accelerators.

The key contributions of this paper are:
\begin{itemize}
    \item A compact, efficient \gls{pe} leveraging the \glspl{fpu}' pipeline registers for buffering.
    \item A scalable, semi-systolic outer-product architecture that sustains 1\,GHz at 0.72\,V on a 12\,nm node across multiple engine sizes and floating-point datatype support, with single percentage digit buffering overhead and above 99\% \gls{fpu} utilization.
    \item A \gls{ppa} evaluation of a $16 \times 16$ \texttt{FP16}-based O-POPE instance demonstrating {\PerfGain} higher GFLOPS, {\AreaGain} higher GFLOPS$/mm^2$, and {\PowerGain} higher TFLOPS$/W$ compared to the best reported $16 \times 16$ \texttt{FP16}-based state-of-the-art \gls{gemm} accelerators in configurations with the same number of \gls{mac} units.
\end{itemize}
\section{Architecture} 
\label{sec:03_architecture}
Figure \ref{fig:architecture} describes the O-POPE architecture, the O-POPE \gls{pe}, the \gls{gemm} execution flow, and the \gls{fpu} inputs and outputs sequence.
% \gls{pe}
\subsection{O-POPE \acrlong{pe}}
The O-POPE \acrlong{pe} comprises two configurable modules: a \gls{mac} unit and design-time configurable $q$-bit accumulators. 

O-POPE can integrate different \gls{mac} units, as required by the supported numerical representation.
In this work, we focus on floating-point support and integrate the open-source IEEE-compliant \gls{fpu} FPnew.
The FPnew embeds dedicated \gls{mac} units for the specified datatypes, including mixed-precision dot-product operations.
To fulfill \gls{mac} requirements, the FPnew accepts $3 \times q$ bits in input and generates $q$ bits in output.
When the input requires fewer than $q$ bits (e.g., \texttt{FP8} inputs and \texttt{FP16} output), FPnew exploits SIMD packing.

FPnew can be instantiated with a configurable number of pipeline registers, depending on the targeted frequency. 
The degree of pipelining determines both the operand reuse pattern and the latency of a single \gls{mac} operation, i.e., the minimum number of cycles required before a new accumulation can be issued on the same output element.

For the sake of illustration, in Figure~\ref{fig:architecture}d, we show a high-performance configuration with four pipeline stages in the \gls{fpu}.
In this configuration, the unit consumes two pairs of input operands ($a_{00}$, $a_{01}$, $b_{00}$, $b_{01}$) and produces four rank-1 updates ($c_{00}$, $c_{01}$, $c_{10}$, $c_{11}$), with a reuse of two for each input operand. Note that the number of rank-1 updates matches the latency of the pipeline, hence the first output of the \gls{fpu} ($c_{00}$) is ready exactly when the $a$ and $b$ inputs of a new rank-1 update are ready to be accumulated.
This balance allows pipeline latency (e.g., four cycles) to be fully hidden without introducing input buffer registers, as the \gls{fpu} output can be directly reinserted as an input C operand.

\begin{figure}[t]
    \centering
    \includegraphics[width=0.45\textwidth]{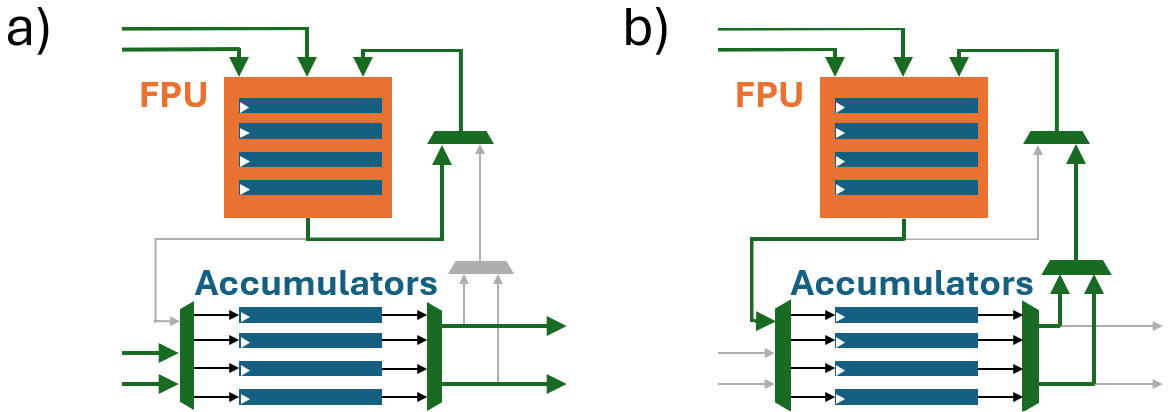}
    \caption{Runtime \gls{fpu}-accumulators decoupling (a) and coupling (b).}
    \label{fig:tile}
\end{figure}

\begin{figure*}[t]
    \centering
    \includegraphics[width=\textwidth]{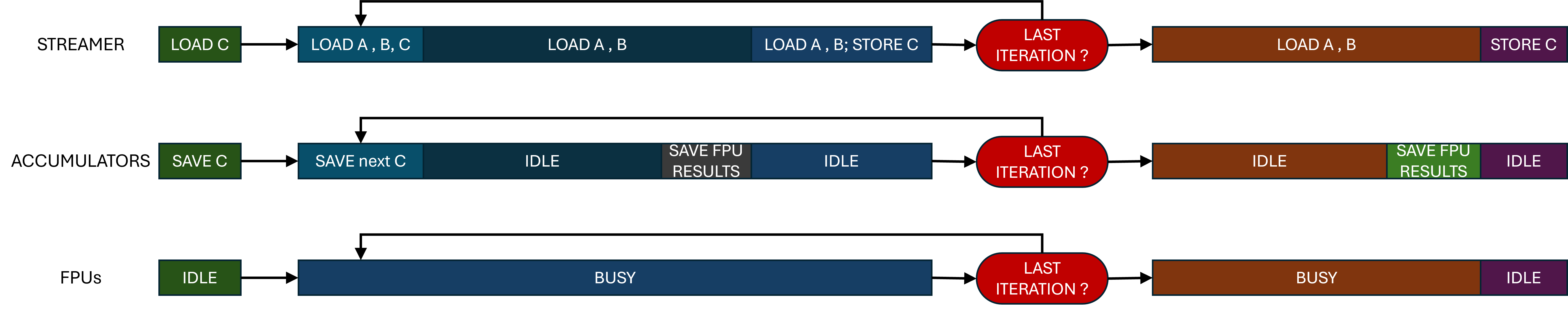}
    \caption{\gls{gemm} execution flow in the streamer, accumulators and \gls{fpu}s.}
    \label{fig:waveforms}
\end{figure*}
However, at the end of the rank-1 updates needed to produce a tile of matrix C (refer to Figure \ref{fig:architecture}c), the four final accumulator values need to be buffered to enable their write-out to external memory. The four accumulator buffer registers at the bottom of Figure \ref{fig:architecture}b have precisely this role. 
By either saving the initial value of the tile or the final \gls{fpu}'s output, these registers allow for decoupling \gls{mac} operations from memory operations.

As shown in Figure \ref{fig:tile}a, when the \gls{fpu} accumulates, the accumulator registers are decoupled and then available to communicate with the memory unit to store a computed value or preload a new initial value.
In contrast, when the \gls{fpu} completes the accumulation, the accumulator registers save the \gls{fpu}'s output and provide the new starting value to the \gls{fpu} without any cycle loss (Figure \ref{fig:tile}b).
For this reason, the number of accumulator registers must match the number of pipeline registers in the \gls{fpu} (four, in the exemplary embodiment of Figure \ref{fig:architecture}).

This compact and configurable \gls{pe} enables high utilization by exploiting the key features of an output-stationary dataflow based on outer-product updates and by leveraging \gls{fpu} pipeline registers as a double buffer.

% Engine
\subsection{O-POPE}
The O-POPE architecture consists of a configurable square mesh ($p \times p$, where $p$ is a power of 2) of the \gls{pe}s described in the previous subsection. 
The engine reads $p \times q$ bits from matrix A and matrix B, broadcasting $q$ bits row-wise for A and column-wise for B.
For matrix C, the engine instead provides a $2p \times q$ bits/cycle input bandwidth to speed up accumulator preloading and a $2p \times q$ bits/cycle output bandwidth to accelerate output writeback to memory.
The propagation of the C elements across engine rows toward the write-out port occurs systolically. This solution favors scalability, as we avoid global connections to the \gls{pe}s and bulky multiplexer structures to read from the engine.

As illustrated in Figure \ref{fig:architecture}d, O-POPE reuses input vectors twice in four consecutive cycles, producing a single update on each accumulator.
Then, the size of each input vector processed for a single accumulator update is $2p \times q$ bits, and the overall output update is $2p \times 1 \times 2p$.
Consequently, O-POPE presents only two $2p \times q$-bit input buffers, which scale with the square root of the number of \gls{pe}s.

In conclusion, our engine ensures scalability and minimal buffer-overhead.

% Compute cluster
\subsection{Integration in a compute cluster}
\begin{figure}[t]
    \centering
    \includegraphics[width=0.45\textwidth]{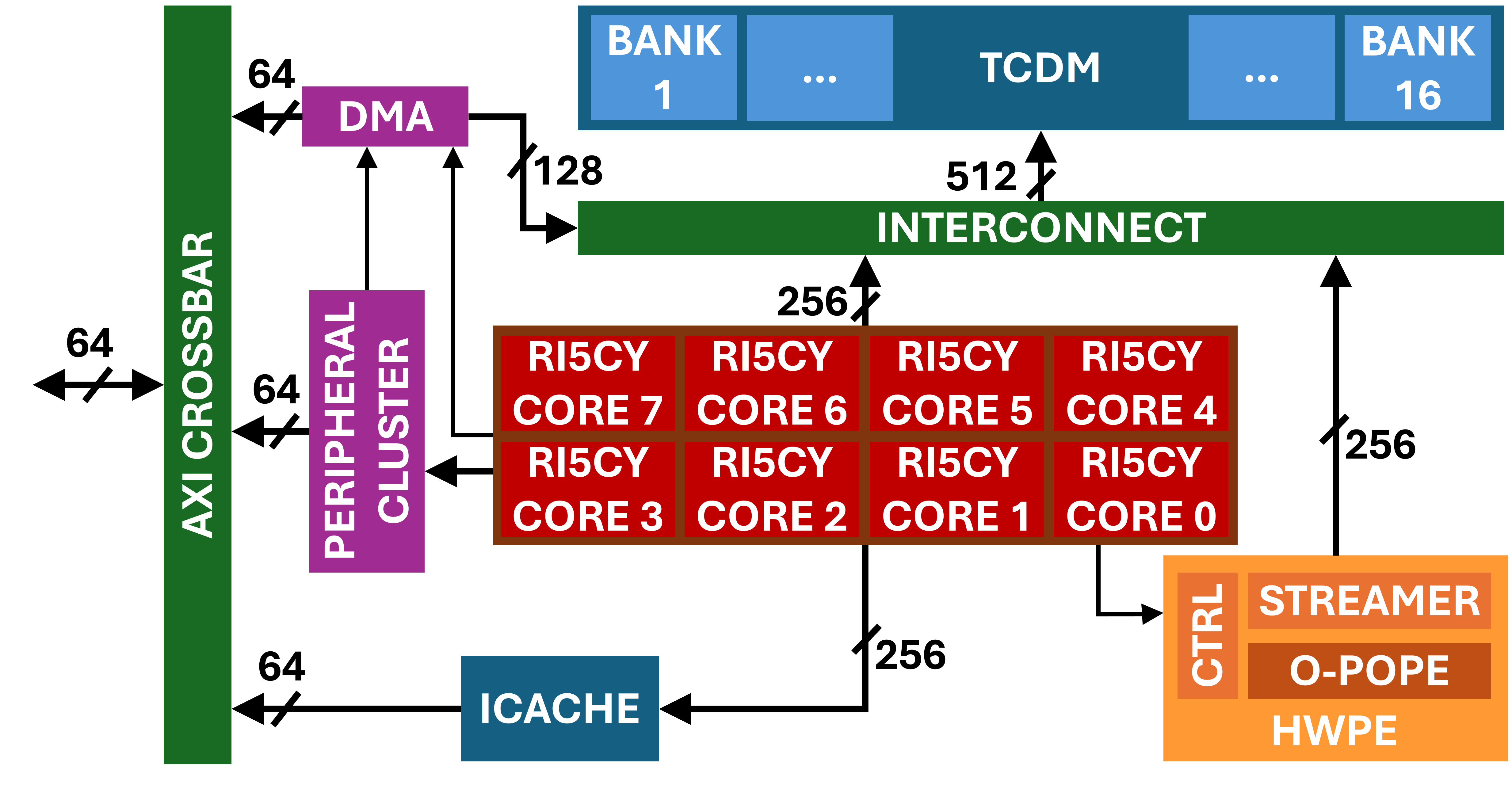}
    \caption{The PULP cluster architecture.}
    \label{fig:pulp_cluster}
\end{figure}

To evaluate and benchmark our design on a broad range of matrix multiplications, we integrate O-POPE within the PULP \cite{rossi2015pulp} cluster via the \gls{hwpe} wrapper \footnote{https://hwpe-doc.rtfd.io}, as shown in Figure \ref{fig:pulp_cluster}. 

Both the cores and O-POPE access a shared \gls{tcdm} of 128kB, through a single-cycle latency interconnect. 
This scratchpad memory is organized in multiple banks to provide enough bandwidth for processing elements' requests.  
To hide the latency of data transfers between L2 and L1 memories, the core $0$ programs the \gls{dma} unit, ensuring double-buffering in the \gls{tcdm}.
An AXI crossbar connects the PULP cluster to the on-chip L2 memory and the host core, which communicates with the PULP cluster via the peripheral cluster.

The \gls{hwpe} wrapper consists of a controller and a streamer that interface the O-POPE engine.
The controller, programmed by the core $0$, orchestrates O-POPE's operation. 
The streamer generates memory addresses according to the sequence illustrated in Figure \ref{fig:waveforms} to transfer O-POPE's input and output data.
To ensure sufficient input bandwidth and full utilization of the \gls{mac} array, we configure the streamer to support transfers of $2 \times p \times q$ bits per cycle.
As mentioned in subsection III.B, the engine requires two $2p \times q$-bit input vectors every four cycles.
Consequently, input vectors utilize 50\% of the bandwidth, with the remaining 50\% left to move the output tile into and out of the engine.
Additionally, we place FIFOs with a design-time-configurable number of reserved slots between the streamer and the engine to enable latency tolerance during transfers from L1 memory in the presence of L1 banking conflicts.

\begin{figure*}[t]
    \centering
    
    \begin{subfigure}[t]{0.45\textwidth}
        \centering
        \includegraphics[width=\linewidth]{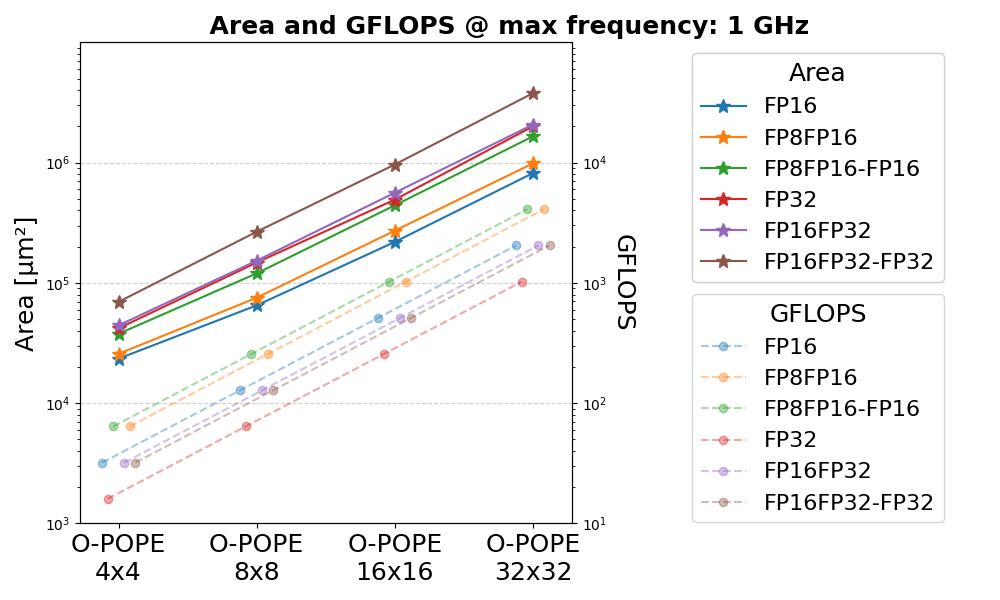}
        \caption{Area and GFLOPS scaling}
        \label{fig:area}
    \end{subfigure}
    \hfill
    \begin{subfigure}[t]{0.35\textwidth}
        \centering
        \includegraphics[width=\linewidth]{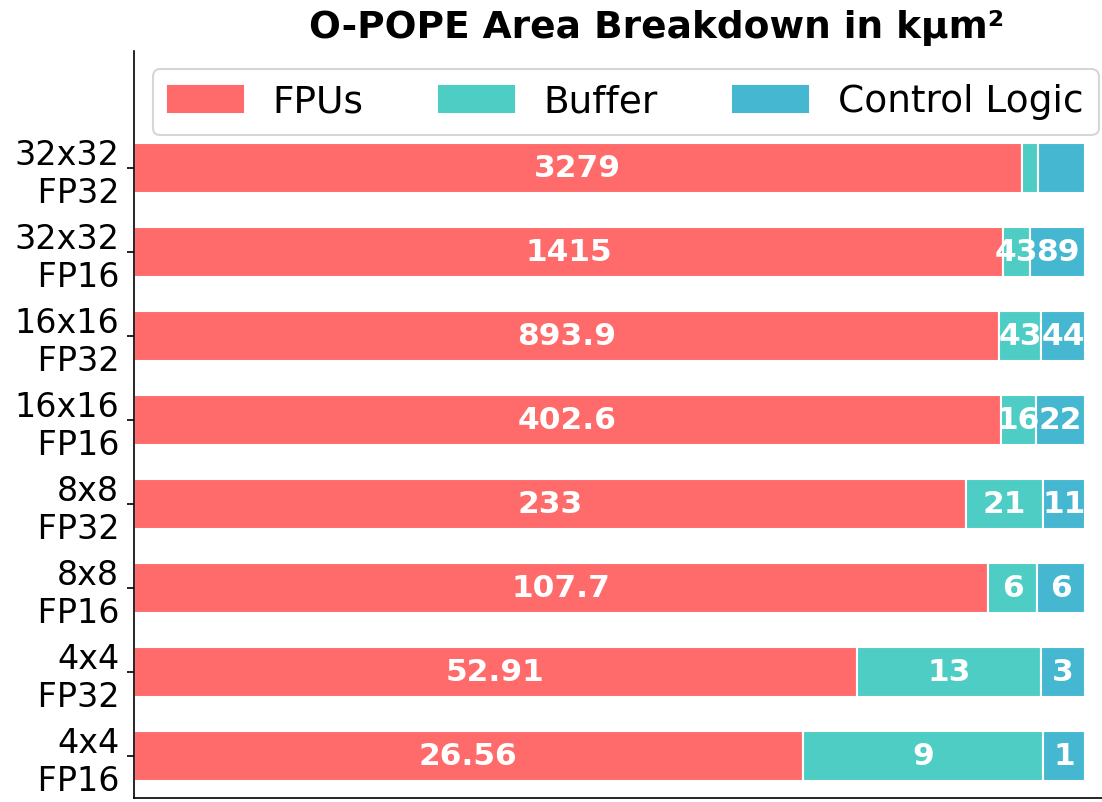}
        \caption{Area breakdown}
        \label{fig:area_bd}
    \end{subfigure}
    \hfill
    \begin{subfigure}[t]{0.14\textwidth}
        \centering
        \includegraphics[width=\linewidth]{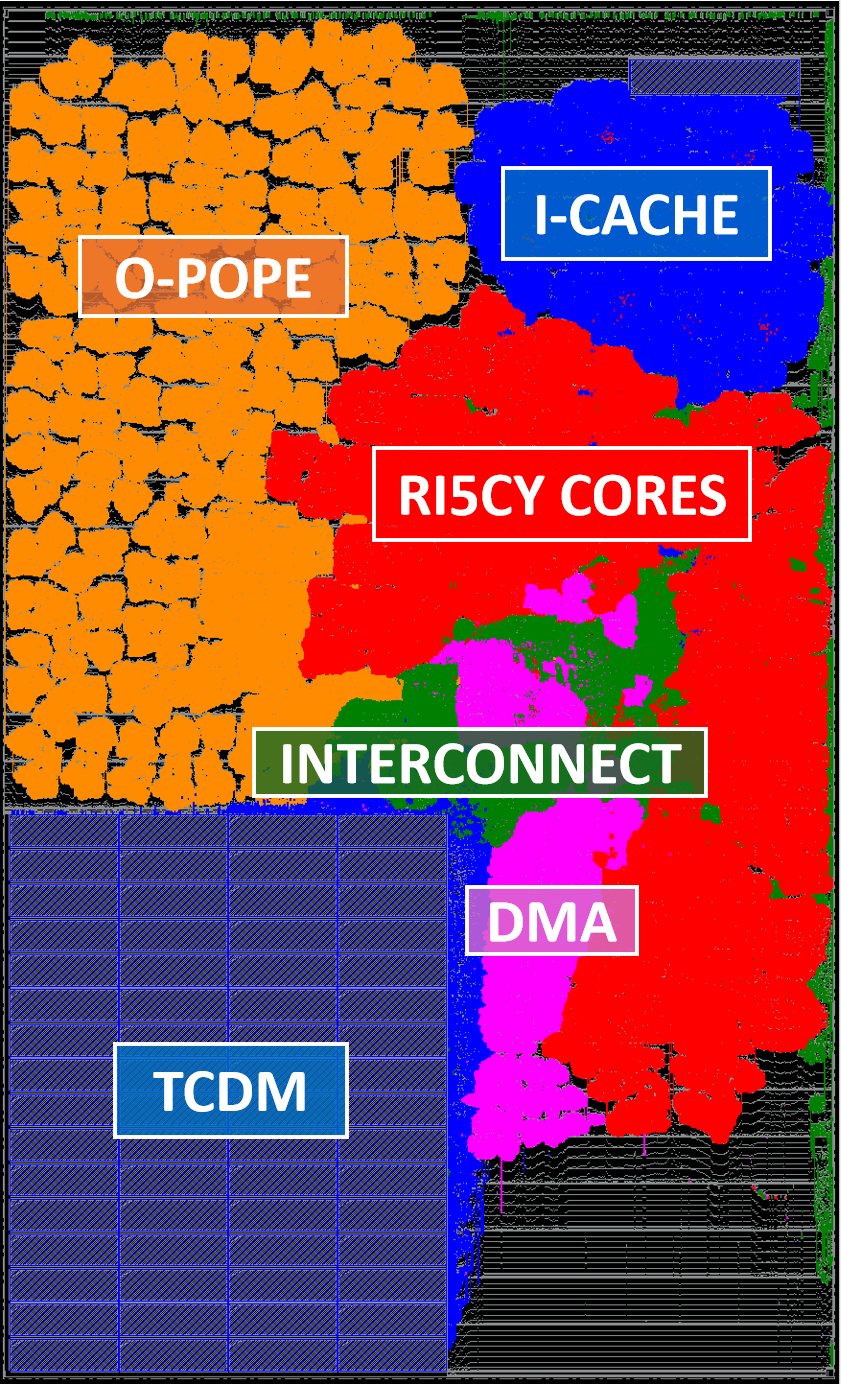}
        \caption{Physical design}
        \label{fig:pd}
    \end{subfigure}

    \caption{O-POPE implementation results.
    (a) Area and GFLOPS scaling at 1 GHz across multiple engine sizes and IEEE-compliant MAC support.
    (b) Area breakdown across compute configurations.
    (c) Physical layout of the O-POPE-enhanced PULP cluster. }
    
    \label{fig:ope_results}
\end{figure*}
The operation of O-POPE in the context of the cluster is illustrated in Figure \ref{fig:waveforms}.
Initially, the streamer loads the accumulators with the initial C values.
Then, it loads a subset of the matrices' rows or columns as indicated in Figure \ref{fig:architecture}c: a vector from matrix A, a vector from matrix B, and two vectors from matrix C in preparation for the forthcoming tile computations. 
The streamer repeats these operations until all accumulators of the forthcoming tile are fully loaded.
Concurrently, the engine starts processing the data and populating the \glspl{fpu}' pipeline registers, as \glspl{fpu} and accumulators are decoupled.
Once the accumulators hold the subsequent C tile's initial values, the streamer fetches vectors exclusively from matrices A and B.
Upon completing the C tile's computation, the \glspl{fpu} and the accumulators couple: the \glspl{fpu} writes the result into the accumulators, and the accumulators feed the \glspl{fpu} with the preloaded C tile. 
From this cycle, the streamer continuously loads a vector from matrix A and one from matrix B and stores two vectors from the computed C tile, while the \glspl{fpu} accumulates.
The process described above repeats until all the C tiles are fetched.
During the final tile computation, the streamer loads vectors from the matrices A and B, then stores the last tile, completing the \gls{gemm} execution.

This controller enables us to exploit the pipeline registers as a double buffer and to efficiently utilize the \gls{fpu}s since they stall only during the preload of the first tile and the writeback of the last tile result.

\section{Experimental Results}

\subsection{Experiment Setup}
To evaluate the \gls{ppa} of our design, as well as its scalability and flexibility, we define the following strategy.

First, we conduct synthesis across several configurations with Synopsys Design Compiler 2021.06 in GlobalFoundries' 12LP+ FinFET technology under worst-case conditions (SS, 0.72~V, 125$^\circ$C) to evaluate the scalability of O-POPE, targeting the same maximum frequency (1\, GHz) achieved by our smallest configuration.

Then, we analyze the utilization of our architecture through cycle-accurate runtime simulations with QuestaSim 2022.3 in a streamlined PULP cluster by executing \gls{gemm} workloads. 
For each of them, we vary the mesh size of our engine from $4\times4$ to $32\times32$, and we sweep several floating-point data types (\texttt{FP8$\rightarrow$FP16}, \texttt{FP16$\rightarrow$FP16}, \texttt{FP16$\rightarrow$FP32}, \texttt{FP32$\rightarrow$FP32}).

% Finally, we perform physical design of the PULP cluster with O-POPE integrated using Cadence Innovus 21.17, and estimate power consumption at 1\, GHz using Synopsys PrimeTime 2022.03 under typical-case conditions (TT, 0.8~V, 25$^\circ$C).
Finally, we physically implement the PULP cluster enhanced with O-POPE (Figure \ref{fig:pd}) using Cadence Innovus 21.17, and estimate power at 1\, GHz using Synopsys PrimeTime 2022.03 under typical-case conditions (TT, 0.8~V, 25$^\circ$C).

\subsection{Area Scalability and Frequency}
Figure~\ref{fig:area} shows post-synthesis area estimations and GFLOPS for O-POPE on a logarithmic scale across different mesh sizes ($4\times4$, $8\times8$, $16\times16$, and $32\times32$) and multiple IEEE-compliant \gls{mac} units. Specifically, we evaluate: 
(i) a \gls{mac} unit performing $2\times$-widening accumulation (\texttt{FP8$\rightarrow$FP16} or \texttt{FP16$\rightarrow$FP32}), 
(ii) a \gls{mac} unit operating at the same precision (\texttt{FP16} or \texttt{FP32}), and 
(iii) a \gls{mac} unit combining both the previous ones (\texttt{FP8$\rightarrow$FP16} and \texttt{FP16}, or \texttt{FP16$\rightarrow$FP32} and \texttt{FP32}).

The O-POPE architecture shows no frequency drop across all evaluated configurations, achieving 1~GHz with linear area scaling. The geometric mean of the area increase between quadratically increasing mesh sizes (e.g., from a $4\times4$ to an $8\times8$ mesh) with the same data support ranges from $3.27\times$ to $3.79\times$, while the peak GFLOPS scale by $4\times$.
The ratio between area scaling and peak performance remains constant: this confirms the absence of implementation bottlenecks when increasing the MAC/cycle across a substantial range.

Figure 6b illustrates the area breakdown of the evaluated engine configurations. We note that a desirable consequence of the outer-product architecture is that the input buffer overhead decreases with respect to the core array area. As a consequence, the area efficiency increases, and for a $32 \times 32$ configuration, the area overhead associated with input buffers drops below 2\%.
\begin{figure}[t]
    \centering
    \includegraphics[width=0.48\textwidth]{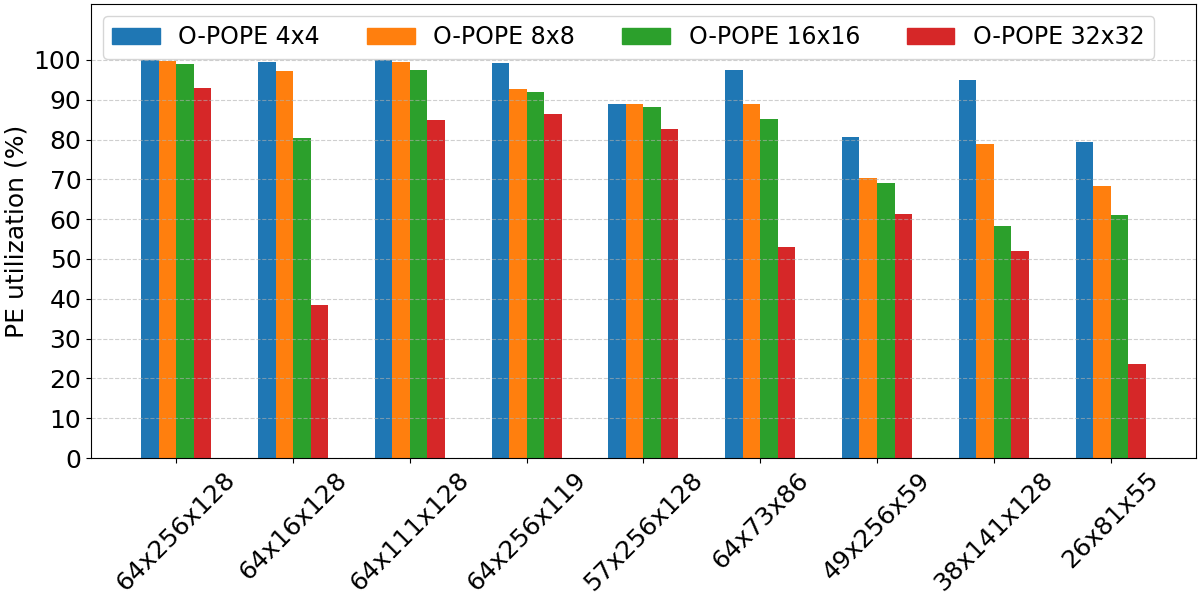}
    \caption{\gls{pe} utilization across arbitrary GEMM sizes.}
    \label{fig:util}
\end{figure}
\subsection{Cycle-Accurate Runtime Analysis}
The runtime analysis results in Figure~\ref{fig:util} report the utilization of \gls{gemm} computations whose input and output matrices fit within the 128~KB \gls{tcdm} of the PULP cluster.\\
Smaller engine dimensions lead to higher utilization because they amortize the initial configuration overhead across a longer execution time. In addition, \gls{gemm} applications whose $M$ and $N$ dimensions are not multiples of twice the engine mesh size (e.g., not multiples of $16$ for an $8\times8$ mesh) experience reduced utilization.
This occurs because O-POPE holds a $4 \times p^2$ $q$-bit output tile to sustain a high operating frequency through four pipeline registers. Hence, iterations that process smaller output tiles underutilize the pipeline and achieve proportionally lower throughput.
Furthermore, we observe lower utilization for small $K$ dimensions, even when $M$ and $N$ are twice the engine mesh size. In such cases, the \gls{mac} operations cannot fully hide the latency associated with moving the output-stationary tile into and out of the engine, as the output tile reuse is too limited when the $K$ dimension is smaller than twice the engine mesh size.
For larger $K$, the utilization increases, as double-buffering in each \gls{pe}, enabled by reusing the \gls{fpu} pipeline registers, allows data movement between the engine and memory to fully overlap with \gls{mac} operations. For \gls{gemm} workloads with both $M$ and $N$ equal to twice a multiple of the engine mesh size and $K$ larger than twice the engine mesh size, O-POPE achieves quasi-ideal utilization, reaching {\MaxUtil} for a $64\times256\times128$ workload in the $4\times4$ configuration.

In case of \gls{ml} workloads, where matrices involved in \gls{gemm} are too large to fit in 128~KB of memory, O-POPE maintains high utilization by employing double buffering in L1 with sufficiently large tiles, as discussed above.
\begin{table}[t]
\centering
\caption{Workload Parameters for M, K, and N}
\label{tab:workloads}
\begin{tabular}{lrrr|lrrr}
\toprule
Workload & M & K & N & Workload & M & K & N \\
\midrule
gMLP\_1 & 196  & 256   & 1536 & ViT\_3      & 197  & 768   & 3072 \\
gMLP\_2 & 196  & 768   & 256  & ConvNext\_0 & 784  & 512   & 256  \\
gMLP\_3 & 768  & 196   & 196  & ConvNext\_1 & 784  & 1024  & 256  \\
MX\_0   & 768  & 196   & 384  & ConvNext\_2 & 3136 & 128   & 512  \\
MX\_1   & 768  & 384   & 196  & ConvNext\_3 & 3136 & 512   & 128  \\
MX\_2   & 196  & 768   & 3072 & GPT3\_0     & 2048 & 768   & 64   \\
MX\_3   & 196  & 3072  & 768  & GPT3\_1     & 2048 & 1024  & 64   \\
ViT\_0  & 197  & 768   & 768  & GPT3\_2     & 2048 & 1536  & 96   \\
ViT\_1  & 197  & 768   & 2304 & GPT3\_3     & 2048 & 128   & 2048 \\
ViT\_2  & 197  & 3072  & 768  &             &      &       &      \\
\bottomrule
\end{tabular}
\end{table}
\begin{figure}[t]
    \centering
    \includegraphics[width=0.48\textwidth]{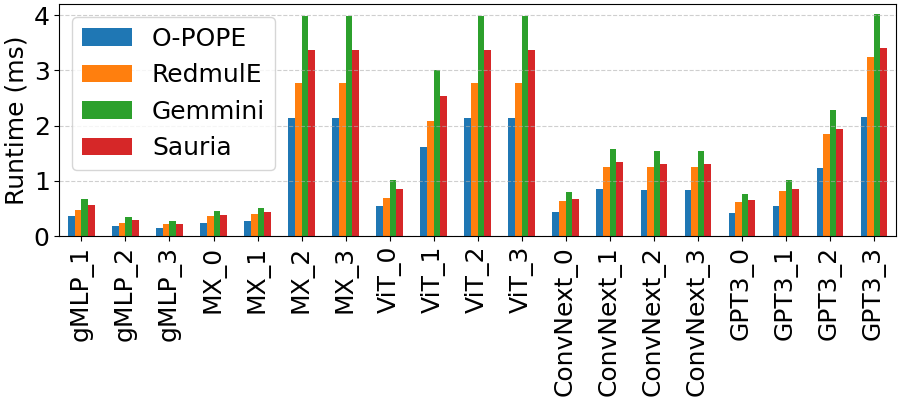}
    \caption{Runtime comparison of $16 \times 16$ floating-point \gls{gemm} accelerators supporting \texttt{FP16 $\rightarrow$ FP16} MAC units on several ML layers.}
    \label{fig:runtime}
\end{figure}

\begin{table}[t]
    \centering
    \caption{Comparison with state-of-the-art $16 \times 16$ \gls{gemm} accelerators supporting \texttt{FP16 $\rightarrow$ FP16} MAC units in 12 nm.}
    \begin{tabular}{cccc}
        \toprule
                & GFLOPS & GFLOPS$/mm^2$ & TFLOPS$/W$ \\
        \midrule
        Gemmini & $280$    & $749$         & $-$        \\
        RedMulE & $384$    & $2134$        & $2.74$     \\
        Sauria$^{\ddagger}$  & $333$    & $1036$        & $2.95$      \\
        \midrule
        \textbf{O-POPE}  & $\textbf{512}$    & $\textbf{2336}$        & $\textbf{3.18}$     \\
        \bottomrule
    \end{tabular}
    \label{tab:soa}
    \\[1pt]
    \footnotesize $^{\ddagger}$ We apply the technology scaling methodology from \cite{sarangi2021deepscaletool}
\end{table}
\subsection{Comparison with State-of-the-Art}

We compare O-POPE against state-of-the-art \gls{gemm} accelerators, namely Gemmini, RedMulE, and Sauria. 

RedMulE and Sauria papers \cite{redmule,sauria2023} analyze in detail a $16\times16$ configuration based on an \texttt{FP16} FPnew instance. To ensure a fair comparison, we adopt the same configuration and floating-point unit for O-POPE and configure Gemmini accordingly.

We evaluate all mentioned accelerators against representative layers from \gls{ml} workloads, including convolutional layers, linear feed-forward layers, and self-attention layers, extracted from widely used models such as Vision Transformers (ViT-Base), convolutional networks (ConvNeXt), Transformer-based language models (GPT), gated MLP architectures (gMLP), and MLP-Mixer.

Figure~\ref{fig:runtime} reports runtime across the \gls{ml} layers listed in Table~\ref{tab:workloads}.
For this evaluation, we integrate O-POPE into the PULP cluster and enable overlap between \gls{dma} transfers and engine execution through double-buffering. Specifically, the DMA operates on 64 KB of \gls{tcdm}, while the remaining 64 KB is allocated to O-POPE.
Core $0$ programs both the DMA and the accelerator using a tiling strategy that satisfies the requirements described in Section IV-C (e.g., 64×128×128 GEMM) and reprograms both components for each tile. This tiling and double-buffering strategy, combined with the increased GFLOPS enabled by pipeline register insertion, allows O-POPE to outperform reported state-of-the-art \gls{gemm} accelerators by up to $1.86\times$.

Table \ref{tab:soa} summarizes GFLOPS, GFLOPS$/mm^2$, and $TFLOPS/W$ for O-POPE and the baseline accelerators in 12-nm technology using $16 \times 16$ \texttt{FP16} FPnew \gls{mac} units. By repurposing \gls{fpu} pipeline registers as double buffers and efficiently supporting output-stationary outer-product execution, O-POPE achieves superior results. Specifically, O-POPE improves GFLOPS by {\PerfGain} over RedMulE, GFLOPS$/mm^2$ by {\AreaGain} over RedMulE, and TFLOPS$/W$ by {\PowerGain} over Sauria.

\section{Conclusion} 
\label{sec:05_conclusion}
O-POPE demonstrates that high-frequency floating-point \gls{gemm} acceleration can be achieved without the buffering overhead traditionally associated with systolic architectures. 
By exploiting FPU pipeline registers as zero-overhead buffers for aligning the accumulation updates with the latency of the \gls{fpu}, O-POPE transforms pipeline latency from a performance constraint into a resource enabling data reuse and synchronization across processing elements. Combined with an output-stationary outer-product dataflow, this approach enables high-frequency operations (1 GHz, 0.72 V) while maintaining extremely high arithmetic utilization.

The proposed semi-systolic architecture scales efficiently across a wide range of \gls{mac} array dimensions and floating-point precisions, preserving performance and frequency and minimizing input buffer overhead. Comparison with state-of-the-art floating-point GEMM accelerators in a 12 nm technology node shows that O-POPE improves performance by {\PerfGain}, area efficiency by {\AreaGain}, and energy efficiency by {\PowerGain}.
% \section*{Acknowledgment}

% The preferred spelling of the word ``acknowledgment'' in America is without 
% an ``e'' after the ``g''. Avoid the stilted expression ``one of us (R. B. 
% G.) thanks $\ldots$''. Instead, try ``R. B. G. thanks$\ldots$''. Put sponsor 
% acknowledgments in the unnumbered footnote on the first page.
\section*{Acknowledgment}
% Hidden for double-blind review.
This work has received funding from the Swiss State Secretariat for Education, Research, and Innovation (SERI) under the SwissChips initiative.
\bibliographystyle{IEEEtran}
\bibliography{src/main}
\end{document}